\documentclass[journal=jacsat,manuscript=article]{achemso}
\setkeys{acs}{articletitle = true}

\usepackage[T1]{fontenc}       

\usepackage[version=3]{mhchem} 
\usepackage{graphicx}
\usepackage{amsmath}
\usepackage{amsthm}
\usepackage{amsfonts}
\usepackage{amssymb, latexsym}
\usepackage{float}
\usepackage{filecontents}

\usepackage[usenames,dvipsnames]{color}
\usepackage[normalem]{ulem}
\author{Mohammad Abo Jabal}
\affiliation[Technion - Israel Institute of Technology]{Wolfson Department of Chemical Engineering, Technion - Israel Institute of Technology, Haifa, Israel 32000}
\author{Ala Egbaria}
\affiliation[Technion - Israel Institute of Technology]{Wolfson Department of Chemical Engineering, Technion - Israel Institute of Technology, Haifa, Israel 32000}
\author{Anna Zigelman}
\affiliation[Technion - Israel Institute of Technology]{Wolfson Department of Chemical Engineering, Technion - Israel Institute of Technology, Haifa, Israel 32000}
\author{Uwe Thiele}
\affiliation[Wilhelms-Universit{\"a}t M{\"u}nster and Center of Nonlinear Science (CeNoS)]{Institut f{\"u}r Theoretische Physik, Westf{\"a}lische Wilhelms-Universit{\"a}t M{\"u}nster, Wilhelm Klemm Str. 9, D-48149 M{\"u}nster, Germany and Center of Nonlinear Science (CeNoS), Westf{\"a}lische Wilhelms Universit{\"a}t M{\"u}nster, Corrensstr. 2, 48149 M{\"u}nster, Germany}
\author{Ofer Manor}
\affiliation[Technion - Israel Institute of Technology]{Wolfson Department of Chemical Engineering, Technion - Israel Institute of Technology, Haifa, Israel 32000}
\email{manoro@technion.ac.il}

\title{Connecting monotonic and oscillatory motions of the meniscus of a volatile polymer solution to the transport of polymer coils and deposit morphology}

\abbreviations{}

\begin{document}

%
%
%
%
%


\begin{abstract}
We study the connection between the polymer deposition patterns to appear following the evaporation of a solution of poly-methyl-methacrylate (PMMA) in toluene, the transport of the polymer coils in the solution, and the motion of the meniscus of the solution. Different deposition patterns are observed when varying the molecular mass and initial concentration of the solute and temperature and are systematically presented in the form of morphological phase diagrams. The modi of deposition and meniscus motion are correlated. They vary with the ratio between the evaporation-driven convective flux and diffusive flux of the polymer coils in the solution. In the case of a diffusion-dominated solute transport, the solution monotonically dewets the solid substrate by evaporation, supporting continuous contact line motion and continuous polymer deposition. However, a convection-dominated transport results in an oscillatory ratcheting dewetting-wetting motion of the contact line with more pronounced dewetting phases. The deposition process is then periodic and produces a stripe pattern. The oscillatory motion of the meniscus differs from the well documented stick-slip motion of the meniscus, observed as well, and is attributed to the opposing influences of evaporation and Marangoni stresses, which alternately dominate the deposition process. 
\end{abstract}

\section{Introduction}
Solutions and suspensions, comprising a volatile solvent and a polymer, macromolecules like collagen or DNA, nanoparticles, or carbon nanotubes were found to produce intriguing one- and two-dimensional deposition patterns at receding three-phase contact lines on solid substrate where the motion is induced by solvent evaporation. Examples include
deposits that consist of one or multiple rings (``coffee ring effect'')~\cite{18Deegan:1997fk,17Deegan:2000tw},  branched patterns resulting from fingering instabilities of the receding contact line~\cite{Thiele1998Dresden-phd,75Pauliac}, and regular periodic stripe patterns~\cite{Lin:2010,61Mhatre}. Moreover, pattern deposition in a restricted geometry has been introduced as a simple preparation route for the fabrication of microscopic structures of high fidelity and regularity.~\cite{Byun,41Xu2006}
The deposited patterns find applications in sensing~\cite{1Weissman, 2Xia, 3Ekhlas, 4Asher, 5Velev, 6Lu}, data storage,~\cite{7Sun, 8Pham} and photonic band gap materials~\cite{9Joannpoulos, 10tetreault} as well as in the fabrication of nanostructured templates~\cite{11Aizpurua, 12SunCH} and of ordered porous materials~\cite{13Velev,14Holland, 15Park, 16Braun}. 
\par
The morphology of the deposits is a consequence of the concurrent interplay of several mechanisms, including the flow in the liquid which is generated by the evaporation of the solvent, Marangoni-force and capillarity induced flows~\cite{31Hu, 26AboJabal, 73Fra, 29Fanton, 27Bodiguel, 28Thiele, 32Weon}, etc. 
 Furthermore, the evaporation of a macroscopic droplet or an unbounded film into an open vapor atmosphere usually yields rather irregular patterns of rugged rings or lines~\cite{35Maheshwari, 36Kajiya, 37Gorand, 38Han, 39Byun}. However, restraining the liquid, using physical confinement, is known to yield well organized deposits~\cite{38Han}.  Examples for such confinements are the square-pyramid/wedge-on-flat \cite{39Byun}, sphere-on-flat~\cite{40Hong, 41Xu2006, 42Xu2007}, ring-on-flat~\cite{43Denkov, 44Pauliac} and cylinder-on-flat \cite{45Kwon} geometries as well as two parallel \cite{27Bodiguel,46Leng, 47Li} or tilted plates, crossed cylinders~\cite{48Zeng}, and capillary tubes~\cite{49Abkarian}. Recent reviews of different aspects of these systems are given elsewhere. \cite{38Han,28Thiele,34Larson}
\par
There are few studies that analyze the properties of regular patterns, which are deposited in confined geometries.
Yabu et al.~\cite{50bYabu} employed a computer-controlled apparatus comprising two precisely manipulated sliding parallel glass plates. A polymer solution (polystyrene dissolved in chloroform) was continuously fed into the gap (200$\mathrm{\mu m}$) between the two plates while the upper plate slid over the lower one. The polymer solution formed a meniscus that moved along with the edge of the upper plate. Various types of polymer patterns were left behind the meniscus on the bottom plate following the evaporation of the solvent. Increasing the polymer concentration, three different modi of contact line motion were found, namely, transversally invariant (i.e., along the contact line) continuous receding motion, transversally invariant stick-slip motion, and continuous receding of a contact line with fingering instability. These modi of contact line motion gave rise to three respective types of micrometer-scale patterns, namely, rhombic dot patterns, stripe patterns, and ladder-like patterns of orthogonal sets of lines. Moreover, changes in the sliding speed of the plate influenced the morphology of the patterns. In another work, Hong et al.~\cite{57Hong2007Zou} evaporated a polymer solution (polystyrene dissolved in toluene) in a capillary structure formed by a sphere-on-flat geometry. They found that the variation in the molecular weight, $\mathrm{M}_{\mathrm{w}}$, of polymers can lead to a pronounced change in the morphology of the deposit. At low $\mathrm{M}_{\mathrm{w}}$ values, the evaporative dewetting process left behind randomly distributed dots.  At intermediate $\mathrm{M}_{\mathrm{w}}$ values, concentric rings were formed by a cyclic deposition-recession process, while high $\mathrm{M}_{\mathrm{w}}$ produced concentric rings, rings with fingers, and punch-hole-like structures. In addition, Xu. et al.~\cite{41Xu2006}, investigated the effect of polymer concentration on the morphology of the concentric rings pattern obtained by evaporating a solution of poly[2-methoxy-5-(2-ethylhexyloxy)-1,4-phenylenevinylene] in toluene in a sphere-on-flat geometry. They found that the distance between adjacent rings and their thickness increased with increasing the concentration of the polymer solution.


%
%
%
%

In the present work, we investigate the physical processes that are responsible for qualitative changes in the pattern deposition of a polymer from a volatile solution, and discuss the corresponding changes in contact line motion. The paper is structured as follows.
In ``Experimental Methods'' we describe our experimental system and procedures of measurement and analysis. In ``Experimental Results'' we present different deposition patterns and morphologies of the polymer and corresponding morphological phase diagrams. We discuss in detail the variations in the morphology of the deposits and connect them to changes in the modus of contact line motion. In ``Discussion'' we highlight the interplay of the diffusive and advective transport of the polymer chains in the volatile solution and its contribution to the bifurcations occurring in the deposition process, and in ``Conclusion'' we conclude our findings.

\section{Experimental Methods}\label{exp}
\begin{figure}[H]
    \centering
  \includegraphics[width=6.5 in]{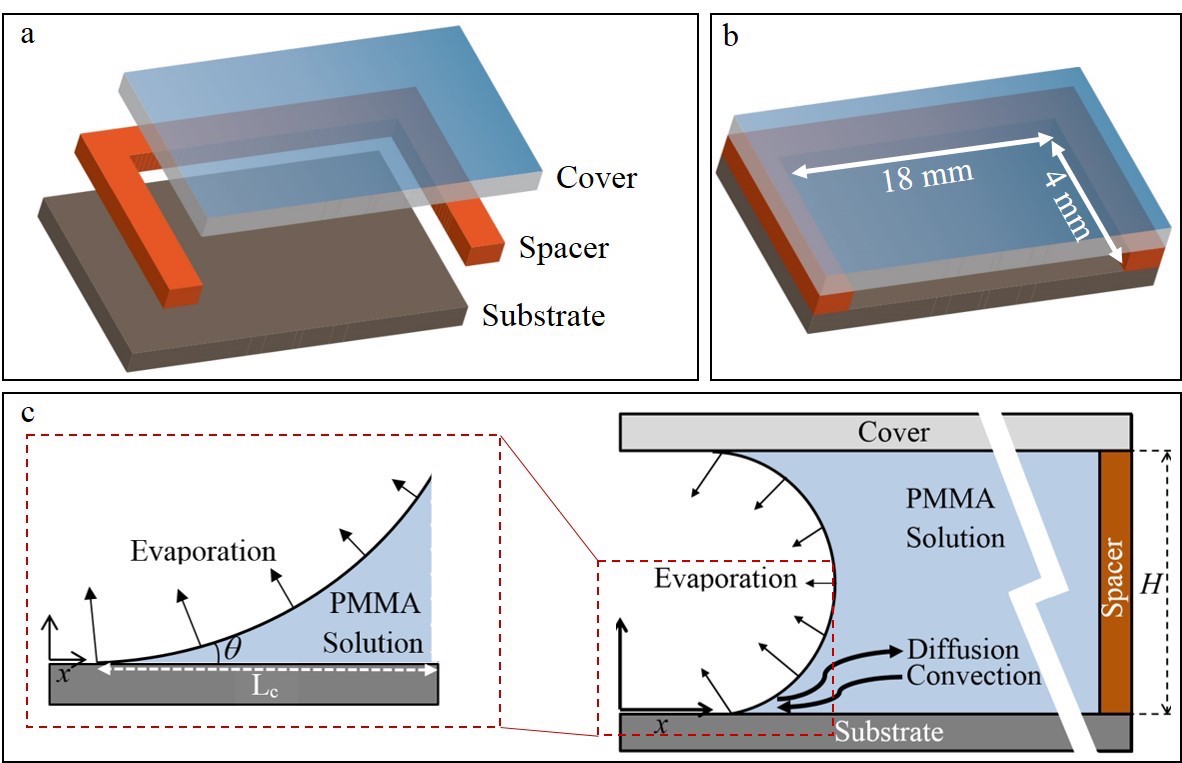}
  \caption{(a) A sketch of the different parts of the micro-chamber shown in (b) and (c) an illustration of the cross-sectional profile of the evaporating solution in the micro-chamber, where $\theta$ and $H$ are the contact angles of the solution with the solid substrate and the inner width of the chamber, respectively.}\label{F:f1}
\end{figure}

As a model experimental system we place a solution of poly-methyl methacrylate (PMMA, 9011-14-7, Aldrich) in toluene ($99.8\%$, 108-88-3, Aldrich) in a micro-chamber of rectangular geometry as illustrated in Fig.~\ref{F:f1}. The chamber is composed of an upper glass cover (microscope slide, Marienfeld), a U-shaped spacer made of polyamide (Kapton, DuPont), and a substrate of silicon oxide (atop a silicon wafer <100>, 550 $\mathrm{\mu m}$ thickness, Silicon Materials).
 We employ polymers chains of average molecular masses of  $\mathrm{M}_{\mathrm{w}}$=$10,000,~ 50,000,$ $100,000$, and $350,000$ Da at initial solute concentrations of  $C =0.5,  1,  3,  5,  7,  10,  13$, and $15$~mg/ml. All initial concentrations are below the overlap concentration, i.e., steric and frictional interactions of neighbouring polymer chains are negligible (dilute solutions). We further employ a chamber with an inner width of $H=75\mathrm{\mu m}$.  
\par

Prior to each experiment we clean the three parts of the chamber by dipping them in a hot bath of acetone, ethanol, isopropyl alcohol, followed by drying them in a stream of air, rinsing in Milli-Q water (18 M$\Omega$ cm), and finally drying again. The substrates are then treated for 5~min using a plasma cleaner (PDC-001-HP (115V), Harrick Plasma).

Each surface is characterised by measuring the apparent receding and advancing contact angles using a goniometer (Data Physics; OCA 15Pro). The receding and advancing contact angles measured for toluene on silicon oxide are $\mathrm{10}^{\mathrm{o}}\pm2^{\mathrm{o}}$ and $\mathrm{12}^{\mathrm{o}}\pm2^{\mathrm{o}}$, respectively, while for toluene on a glass cover these angles are measured to be $\mathrm{14}^{\mathrm{o}}\pm2^{\mathrm{o}}$ and $\mathrm{19}^{\mathrm{o}}\pm2^{\mathrm{o}}$, respectively. The micro-chamber is then placed on a heating plate inside a laminar fume hood (Optimal 12, St\'{e}rile Laminaire).
 As the liquid evaporates, the three-phase contact line between the toluene, its vapor, and the silicon oxide substrate moves (on average) toward the interior of the chamber. The motion of the contact line, the deposition process, and the morphology of the polymer deposits are recorded by a monochromatic camera (Q Imaging, Nikon) mounted on an upright Nikon eclipse optical microscope. Moreover, our measurements of the dynamic position of the contact line further give the 
temporal rate of evaporation of the meniscus in the chamber versus the distance between the contact line and the open end of the chamber.

  

\section{Experimental Results}
\begin{figure}[h!]
    \centering
  \includegraphics[width=6.5 in]{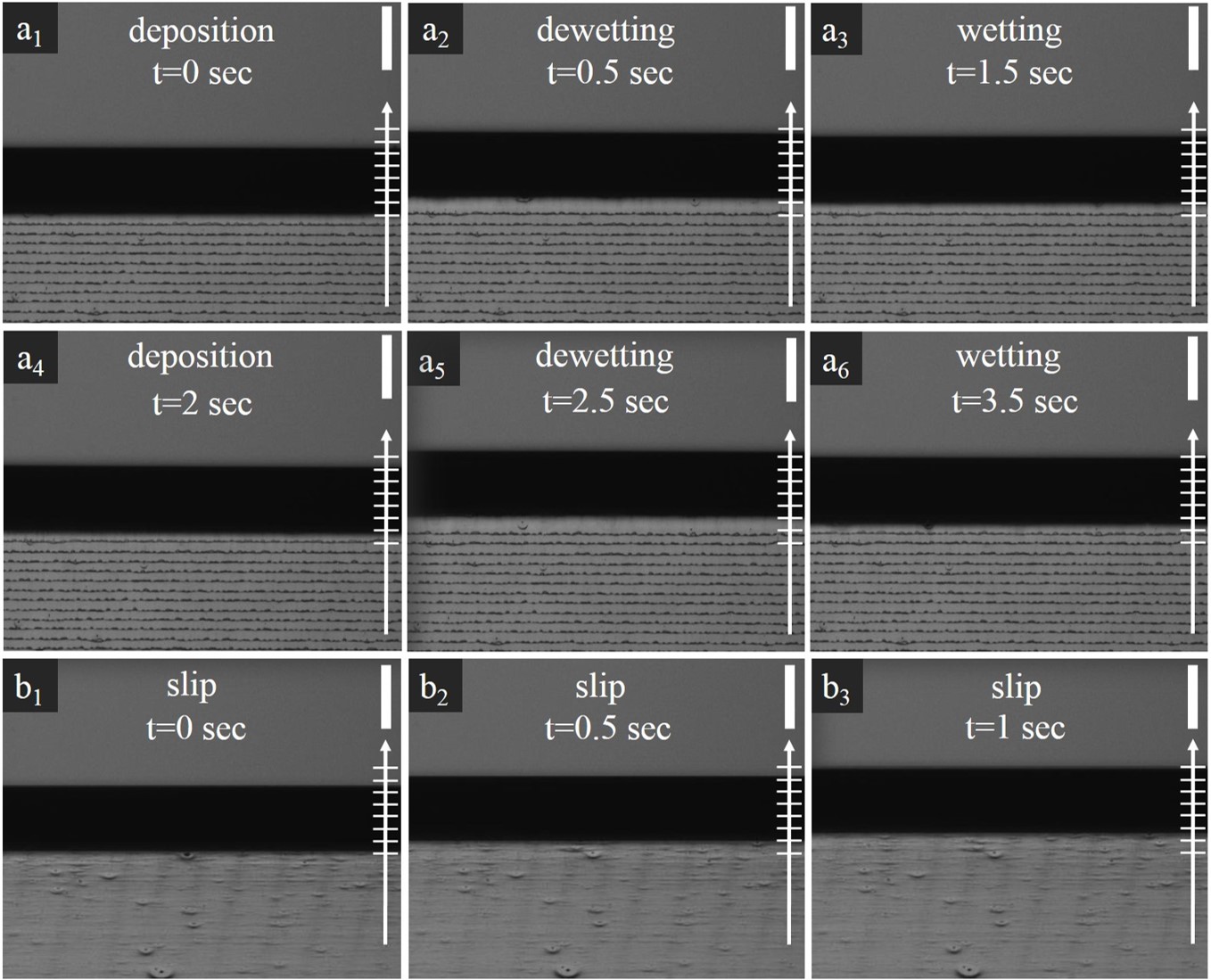}\\
  \caption{Three sequences of video frames (top views) that show time evolutions of the contact line region (meniscus position) and the resulting deposition patterns (lower part of each image) in the course of the evaporative dewetting in the micro-chamber. The large black rectangle in the middle of each image is the curved meniscus. The homogeneously gray parts above it correspond to the part of the chamber which contains the solution. The polymer deposit in the lower part of each frame is nearly homogeneous in the bottom row ($b_{1}$-$b_{3}$), and shows a regular stripe pattern in the top and middle rows ($a_{1}$-$a_{6}$). The corresponding dependencies of the contact line position on time are given by the black and red solid lines in Fig.~\ref{F:f4}, respectively. The only varied parameter is the molecular mass of the polymer that is $M_w=350,000$~Da in ($a_{1}$-$a_{6}$) and $M_w=50,000$~Da in ($b_1$-$b_3$). The open end of the micro-chamber is located at the bottom of each image, the white scale bar is 50 $\mu$m, and the separation between the ticks on the white arrow that indicates the direction of overall contact line motion is 10 $\mu$m. The inner width of the chamber is $H=75$ $\mu$m, the temperature is $T=25^{\mathrm{o}}{\mathrm{C}}$, and the initial solution concentration is $C=3$ mg/ml. }\label{F:f1_4}
\end{figure}

\begin{figure}[h!]
    \centering
  \includegraphics[width=6.5 in]{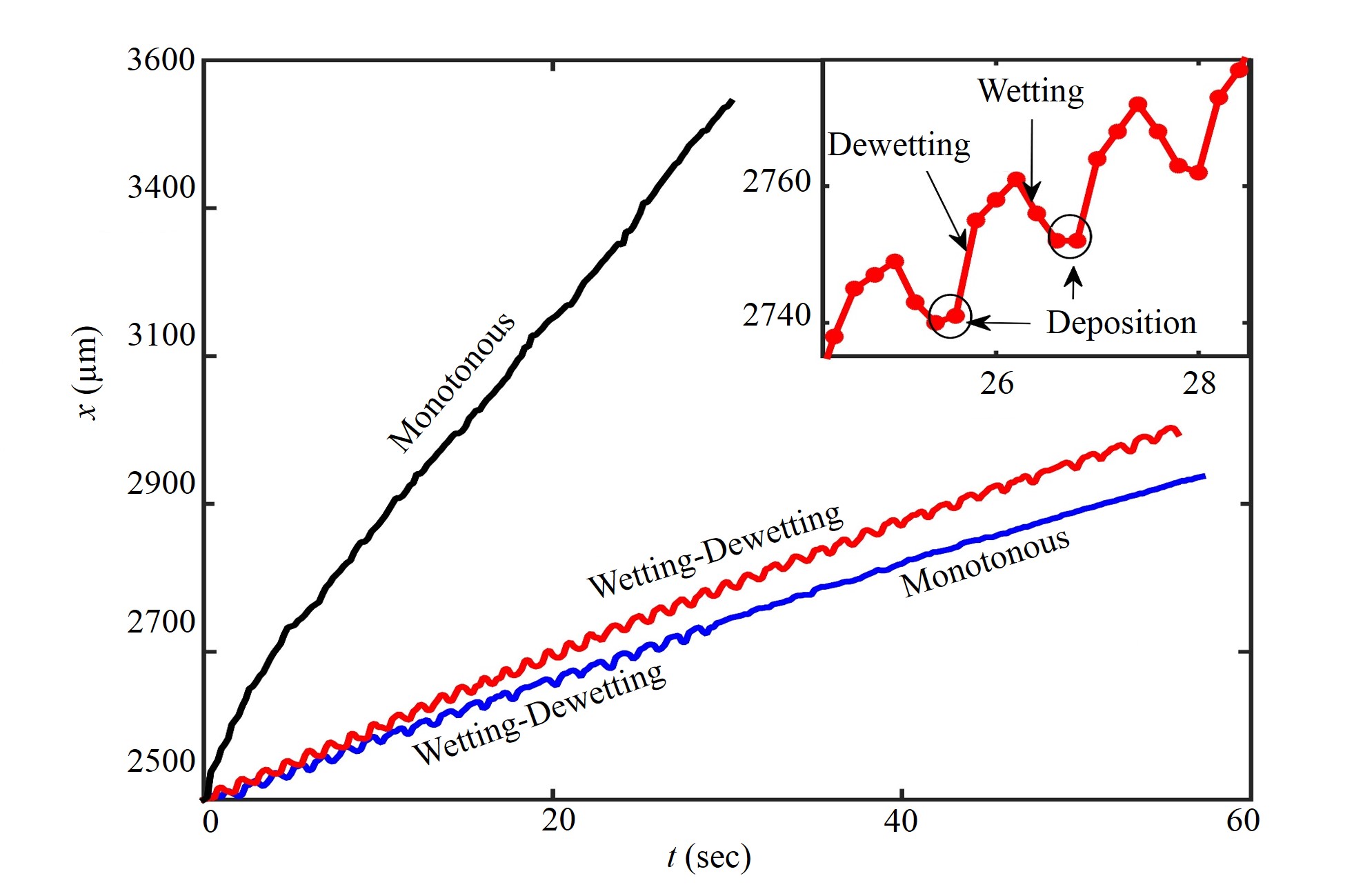}\\
  \caption{Time ($t$) variations of the contact line position ($x$) for the different modi of contact line motion in a micro-chamber, given by a monotonous motion at a molecular mass of $M_w$=50,000 Da (black line), an oscillatory wetting-dewetting motion at $M_w$=350,000 Da (red line), and a transition from oscillatory to monotonous mottions at $M_w$=100,000 Da (blue line) at a temperature of $T=25^{\mathrm{o}}{\mathrm{C}}$, and with a solution concentration of $C=3$ mg/ml, where the open end of the chamber is at $x=0$. The inset gives an enlarged view of a few periods of the oscillatory wetting-dewetting motion.}
\label{F:f4}
\end{figure}

\begin{figure}[h!]
    \centering
  \includegraphics[width=6.5 in]{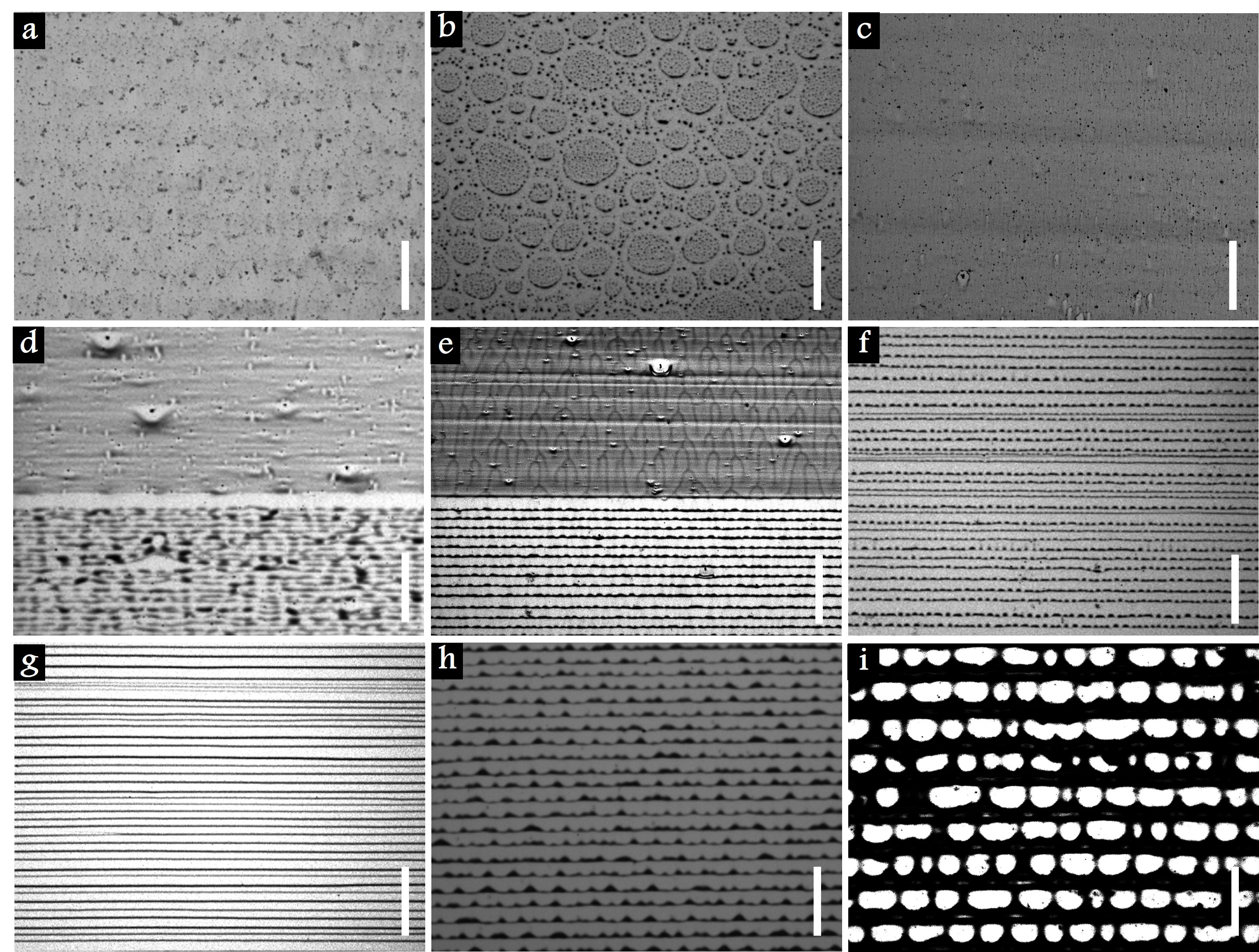}\\
  \caption{Typical deposition patterns, where we show in (a) a non-uniform coating ($M_w=10,000$ Da, $T=25^{\mathrm{o}}{\mathrm{C}}$, $C=0.5$ mg/ml), in (b) a polygonal network pattern ($M_w=10,000$ Da, $T=25^{\mathrm{o}}{\mathrm{C}}$, $C=3$ mg/ml), in (c) a uniform coating ($M_w=50,000$ Da, $T=25^{\mathrm{o}}{\mathrm{C}}$, $C=3$ mg/ml), in (d) discontinuous stripes followed by uniform coating ($M_w=100,000$ Da, $T=25^{\mathrm{o}}{\mathrm{C}}$, $C=3$ mg/ml), in (e) continuous stripes followed by a dendrite-like pattern in an otherwise uniform coating ($M_w=100,000$ Da, $T=25^{\mathrm{o}}{\mathrm{C}}$, $C=7$ mg/ml), in (f) discontinuous stripes  ($M_w=100,000$ Da, $T=50^{\mathrm{o}}{\mathrm{C}}$, $C=7$ mg/ml), in (g) continuous stripes, i.e., a typical ``stripe pattern,'' ($M_w=350,000$ Da, $T=25^{\mathrm{o}}{\mathrm{C}}$, $C=3$ mg/ml), in (h) continuous stripes with moderate fingering instability ($M_w=350,000$ Da, $T=35^{\mathrm{o}}{\mathrm{C}}$, $C=3$ mg/ml), and in (i) a more extreme fingering instability termed ``punch-hole-like'' ($M_w=350,000$ Da, $T=65^{\mathrm{o}}{\mathrm{C}}$, $C=7$ mg/ml). The open end of the micro-chamber is located at the bottom of each image and the white scale bar is 100 $\mu$m.}
 \label{F:f3}
\end{figure}

We distinguish between three characteristic regimes of contact line motion (see movie SI1 in Supplemental Information).
In the first regime, we observe that the contact line moves continuously and monotonically while its distance from the open end of the chamber increases due to evaporation. This happens when employing polymers of low molecular mass, namely, $M_w$=10,000 Da and $M_w$=50,000 Da, at low temperatures and polymer concentrations. The position of the contact line versus time and the deposit that it leaves behind are visualized in the sequence of video frames shown in Fig.~\ref{F:f1_4}$(b_1-b_3)$. The corresponding dependence of the position of the contact line on time is shown in Fig. \ref{F:f4} as a black solid line. The process results in a continuous deposition of either uniform coatings or of continuous structures in the form of polygonal networks; both are shown in Fig.~\ref{F:f3}(a-c). 

The second regime appears when employing the polymer of our largest molecular mass ($M_w$=350,000 Da). Here we observe a different type of contact line dynamics as it undergoes a non-monotonous oscillatory wetting-dewetting motion cycles that are correlated to the deposition of lines. A sequence of video frames visualizing the displacement of the contact line during two cycles and the resulting deposit is shown in Fig.~\ref{F:f1_4}{($a_{1}$-$a_{6}$)}, while the red solid curve in Fig.~\ref{F:f4} gives the contact line position over time and clearly indicates a nearly periodic change of the contact line speed. Specifically, each cycle is composed of three stages. 
Initially, the three-phase contact line recedes for a finite time, i.e., the solution dewets the substrate due to solvent evaporation. This stage is then followed by the motion of the contact line into the opposite direction for a finite time, i.e., the solution wets the substrate again. During these two steps the polymer accumulates close to the contact line. Next, the solution recedes again (dewets the solid) while, leaving a stripe of polymer behind. Moreover, the distance that the contact line traverses during the dewetting stage is greater than during the wetting stage. The combined displacements of the contact line in each cycle result in an overall dewetting motion, which is well visualized in Fig.~\ref{F:f4}. We refer to this modus of contact line motion as an ``oscillatory wetting-dewetting motion''. The
corresponding deposits correspond to discontinuous stripes, solid continuous stripes, and continuous stripes with moderate or extreme fingering instabilities, as shown in Fig.~\ref{F:f3}(f-i)

In the case of moderate values of the molecular mass, namely, $M_w$=50,000 Da (at relatively high temperatures and polymer concentration in the solution) and $M_w$=100,000 Da, we initially observe an oscillatory wetting-dewetting motion of the contact line that later transforms into a monotonic motion when the meniscus has receded deeper into the chamber, shown in Fig.~\ref{F:f4} as a blue solid line. The patterns which correspond to this regime are continuous and discontinuous stripes that are then replaced by a uniform polymer film as shown in Fig.~\ref{F:f3}(d,e).

\begin{figure}[H]
    \centering
  \includegraphics[width=6.5 in]{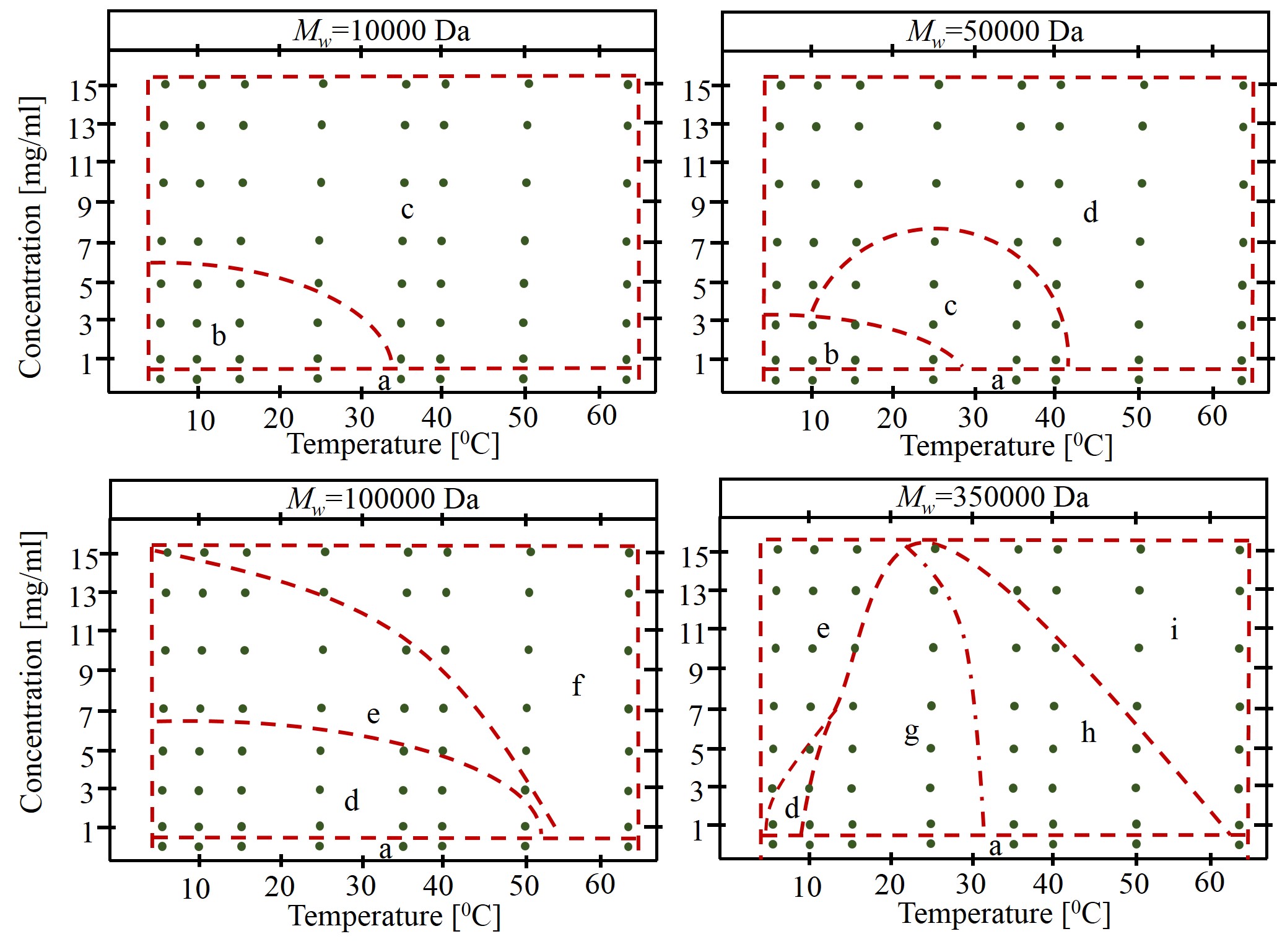}\\
  \caption{Morphological phase diagrams: Temperature and concentration variations of the deposition patterns ``a'' to ``i'' in Fig.~\ref{F:f3} for different polymer molecular masses, $M_w$, where the approximate boundaries between different deposition modi are marked as dashed red lines and each green dot represents an experimentally obtained result.}
  \label{F:f2}
\end{figure}

The state of deposition was generated by choosing specific combinations of the control variables, namely, temperature, polymer concentration, and molecular mass. The different deposition patterns are displayed in Fig.~\ref{F:f3}. The letters ``a''-``i'' that indicate the individual states of deposition correspond to regions in the morphological phase diagrams in Fig.~\ref{F:f2}. 
In Fig.~\ref{F:f3}(a) we show a non-uniform coating that occurs in the regions denoted by ``a'' in the panels in Fig.~\ref{F:f2}. Essentially, this region is characterized by low initial polymer concentrations (below 0.5 mg/ml). At low molecular mass ($M_w=10,000, 50,000$ Da), low temperature, and low concentration, the deposits take the shape of an continuous polygonal network~\cite{79nguyen2002patterning}, as shown in Fig.~\ref{F:f3}(b) and marked by ``b'' in Fig.~\ref{F:f2}. Increasing the temperature and concentration results in the deposition of a uniform film (shown in  Fig.~\ref{F:f3}(c) and marked by ``c'' in Fig.~\ref{F:f2}). Deposits that show transients from discontinuous stripes to uniform deposition are shown in Fig.~\ref{F:f3}(d) and result in the regions denoted by ``d'' in Fig.~\ref{F:f2}. At larger molecular mass ($M_w=100,000$ Da), we additionally observe a transition from the transient patterns marked by ``d'' in Fig.~\ref{F:f2} to the deposition modus shown in Fig.~\ref{F:f3}(e) and marked by ``e'' in the middle row in Fig.~\ref{F:f2}. The deposits change from continuous stripes to a dendrite-like pattern~\cite{17Deegan:2000tw}. Further increasing the temperature and the initial concentration brings us to region ``f'' in Fig.~\ref{F:f2} (see pattern in Fig.~\ref{F:f3}(f)) where discontinuous stripe patterns are found.
Finally, a further increase in the molecular mass (reaching $M_w=350,000$ Da) at low temperature and concentration gives again the patterns in ``d'' in Fig.~\ref{F:f2}. Increasing the initial concentration, but keeping the temperature fixed, gives the deposits ``e'' in Fig.~\ref{F:f2}. However, increasing the temperature for concentrations above 1 mg/ml, we observe an additional type of pattern, namely, a regular continuous stripe pattern (Fig.~\ref{F:f3}(g), marked ``g'' in the lower right panel of Fig.~\ref{F:f2}), which becomes punctured by fingering instability (Fig.~\ref{F:f3}(h), marked ``h'' in the lower right panel of Fig.~\ref{F:f2})~\cite{75bmaillard2001rings}, and the ``punch-hole-like'' pattern (Fig.~\ref{F:f3}(i)) marked by ``i''~\cite{40Hong}.

\section{Discussion}\label{disc}
\subsection{Modi of deposition}
Next we discuss the physical mechanisms which determine the motion of the contact line and the corresponding modi of continuous or patterned deposition of the polymer from the volatile solution. The dynamics of the contact line appears to be connected to the ratio between the rates of solvent evaporation-driven convection of the solution and of polymer diffusion within the solvent. This is captured by the P\'{e}clet number, $\mathrm{Pe}$, for polymer transport~\cite{72Zigelman}, 
\begin{equation}\label{E:eq 1}    
    \mathrm{Pe}\equiv\frac{L_{c}\varepsilon_{\mathrm{ev}}}{D_{l}\theta}.
\end{equation}
Here, $D_l$, $L_c$, $\theta$, and $\varepsilon_{\mathrm{ev}}$ are the diffusion coefficient of the solute, the characteristic length scale of the meniscus (shown in Fig.~\ref{F:f1}c), the contact angle of the liquid with the substrate, and the evaporation rate per unit area, respectively; the velocity of the liquid along the solid is approximately $\varepsilon_{\mathrm{ev}}/\theta$. Further, the diffusion coefficient is a function of the molecular mass of the polymer chains, where according to de Gennes\cite{80masaro1999physical}  $D_{l}=KM_{w}^{-a}$ and $a\simeq2$  is the corresponding power for a diluted polymer solution (below the overlap concentration). Hence, 
\begin{equation}
    \mathrm{Pe}\equiv\frac{\varepsilon_{\mathrm{ev}}L_{c}M_{w}^{a}}{K\theta}.
\end{equation} 
The measured evaporation rate is approximately $\varepsilon_{\mathrm{ev}}\approx10^{-7}$ m/s (see the experimental section) at a temperature and an initial concentration of $T=25^{{\mathrm{o}}}{\mathrm{C}}$ and $C=3$ mg/ml, respectively. The contact angle is estimated as the measured receding contact angle, $\theta\approx 0.17~{\mathrm{rad}}$ ($\backsimeq10^{\mathrm{o}}$), and the characteristic length is taken to be the approximate radius of the meniscus, namely, $L_c\approx H/2\theta$. 

\begin{figure}[h!]
    \centering

   \includegraphics[width=4.25 in]{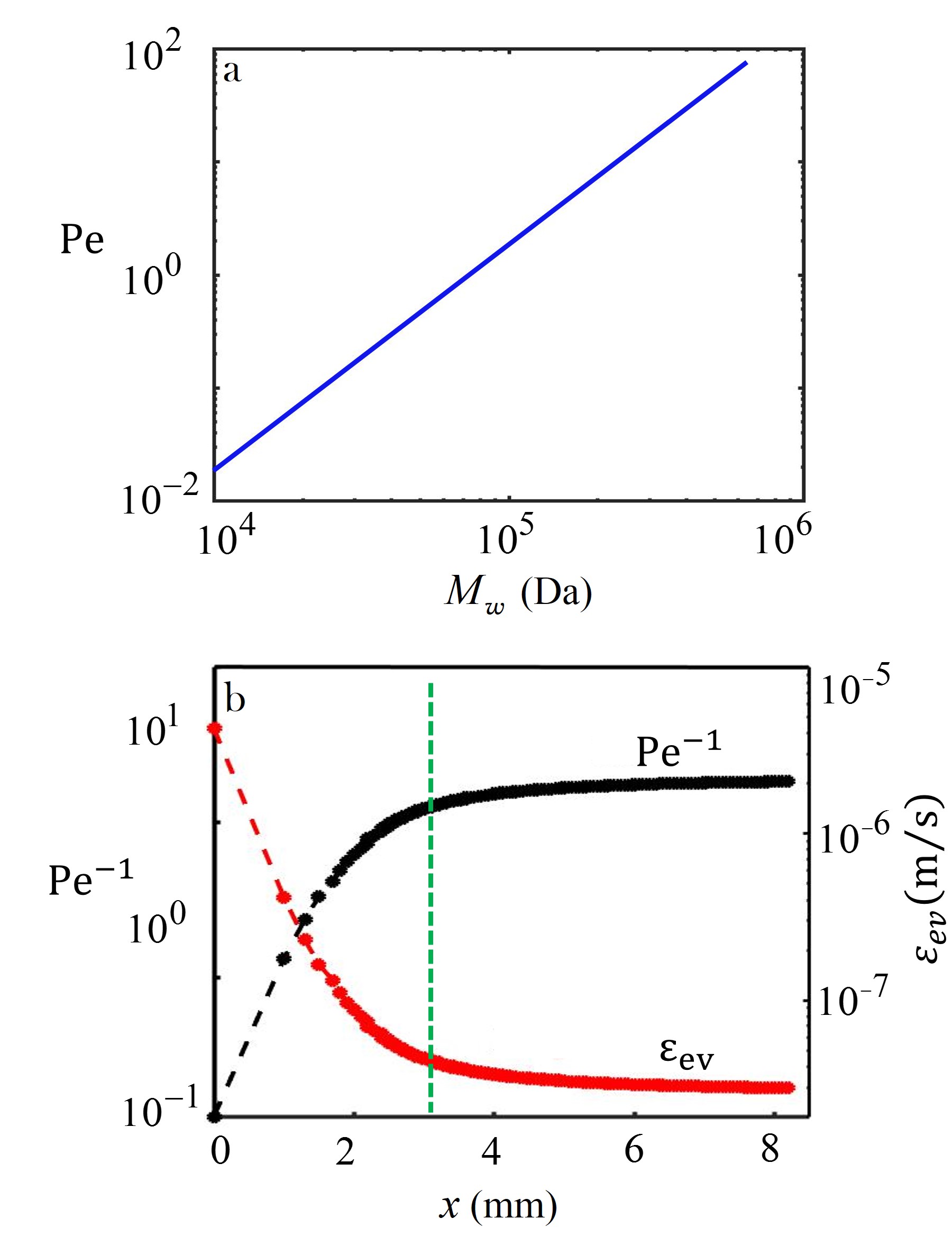}
  \caption{(a) Molecular mass, $M_w$, variations of the P\'{e}clet number, $\mathrm{Pe}$, where we assumed $K\simeq 10^{-1}$ m$^2$ gr$^2$/s mol$^2$ based on the de-Gennes correlation for a given diffusion coefficient of PMMA in Toluene and data given elswhere \cite{BKbrandrup1975polymer}, $T=25^{{\mathrm{o}}}{\mathrm{C}}$, $C=3$ mg/ml, $\theta\simeq 0.17$ rad, $H=100$ $\mathrm{\mu m}$, and $\varepsilon_{\mathrm{ev}}=10^{-7}$ m/s. (b) Separation (contact line to open end of chamber, $x$) variations of the P\'{e}clet number for the same physical parameters as before, excluding $\varepsilon_{\mathrm{ev}}$, and for $M_w=100,000$ Da, where the vertical dashed green line describes the separation where the transitions in the motion of the contact line and in the modus of deposition occur, the black and red points are given at the separations where the rate of evaporation, $\epsilon_{ev}$, was measured, and the dashed black and red lines give the trend of data. }
 \label{F:f9}
\end{figure}

With these values, we show in Fig.~\ref{F:f9} that the P\'{e}clet number increases by nearly two orders of magnitude across the experimental change of the molecular mass of the polymer from $M_w=10,000$~Da to $M_w=350,000$~Da. Thus, when $M_w$ is small, the diffusion of polymer in the solvent may equilibrate the distribution of polymer within the meniscus faster than the evaporation of solvent convects the polymer toward the contact line. As a result, the concentration within the meniscus should be nearly uniform. The deposition of the polymer is then continuous and nearly uniform, as shown in Fig ~\ref{F:f3}(c). The upper left panel in Fig.~\ref{F:f2} shows that this is the prevalent modus of deposition for the lowest investigated molecular mass ($M_w=10,000$~Da) if the concentration and/or temperature are sufficiently high.


The distinction between the non-uniform coating in Fig ~\ref{F:f3}(a) and the polygonal network pattern in Fig ~\ref{F:f3}(b) is likely to result from different dewetting mechanisms of an ultrathin moist or a mushy polymer layer that is left behind the receding contact line. This phenomenon was observed previously in the case of evaporatively dewetting nanoparticle suspensions (see Ref.~\cite{75Pauliac}) and is a reminiscent of the result of a homogeneous thin film of evaporating polymer solution. The nearly continuous patterns in Figs.~\ref{F:f3}(a,b) indicate that the final shape of the deposit is independent of the motion direction of the meniscus. As for evaporating aqueous solutions of collagen, studied by~\citet{MTBL1997saia,ThMP1998prl}, a film with small holes (as in Fig.~\ref{F:f3}(a)) or a polygonal network (as in Fig ~\ref{F:f3}(b)) may result when spinodal dewetting or dewetting via nucleation dominate, respectively. For further detail see the discussions about the two dewetting modes for nonvolatile films \cite{ThVN2001prl,Thie2003epje} and about polygonal network structures, resulting from the dewetting of various volatile and nonvolatile thin films, given elsewhere\cite{TMPW1999,Reit1993l,MoTB2002prl,RoAT2011jpcm} .

At large molecular mass of the polymer the P\'{e}clet number is large, so that the diffusion of the polymer in the solvent is slow when compared to convective flows resulting from evaporation. Hence, the diffusion of polymer coils is not able to equilibrate the polymer concentration within the meniscus. The evaporation of the solvent results in the build up of high polymer concentrations close to the contact line. This results in the deposition of different types of pronounced non-continuous patterns, as shown in Figs.~\ref{F:f3}(f) to (i), and is accompanied by oscillatory wetting-dewetting or stick-slip motions of the receding contact line, shown in Figs.~\ref{F:f4}. 

We now bring our non-continuous deposits into the context of results obtained by Fra{\v{s}}tia et al.,\cite{FrAT2011prl,73Fra,TVAR2009jpm} which modeled related systems by employing thin-film (long-wave) equations for the time evolution of the film height and the local amount of solute \cite{70Oron,67Warner}. They defined a reciprocal P\'{e}clet number, $\mathrm{Pe}^{-1}$, as the ratio between the time scales of the wettability-driven dewetting and solute diffusion. 
%
In particular, they studied the case of a straight evaporatively and convectively receding front of a dewetting (infinite-length) film of nanoparticle suspension and showed that a viscosity that strongly increases with solute concentration is a mechanism for structured deposition. In the regime of line deposition, they described a periodic pinning and depinning dynamics of a receding liquid front to result from a periodic transition between convection-dominated and evaporation-dominated regimes of the front motion.

Fra{\v{s}}tia et al.~\cite{73Fra} further determined the connection between the properties of line patterns, such as length amplitude and period, and various physical parameters. They mainly studied the contributions to the deposition process from evaporation rate and initial solute concentration and described various transitions from uniform to line deposition. Their analysis corresponds to the presently described transitions from the morphology of the deposits given in Figs.~\ref{F:f3}(a) to (c) to the ones in Figs.~\ref{F:f3}(f) to (i). Furthermore, parts of the phase diagrams in Fig.~\ref{F:f2} qualitatively agree with the lower left part of Fig.~9 in Ref.~\cite{73Fra}. Here we take into account that an increase in the evaporation number in the work by Fra{\v{s}}tia et al.\ correlates to an increase in the temperature employed as a control parameter. Fra{\v{s}}tia et al.\ also found that line deposition ceases again at large values of the evaporation number and initial concentrations, which are beyond the parameter values studied here.

The influence of solute diffusion was specifically studied in Section~3.5 of Ref.~\cite{73Fra}. There it was demonstrated that large P\'{e}clet numbers (range $10^{4} - 10^{5}$) support the deposition of stripes. However, reducing the P\'{e}clet number towards $10^{3}$ suppresses stripe deposition as then the receding dewetting front is not able to accumulate the solute, resulting in a uniform coating. This corresponds to the general tendency that we observe when moving from the lower right panel of Fig.~\ref{F:f2} to the upper left one, i.e., decreasing $M_w$ and therefore the P\'{e}clet number. This agreement remains valid even though our present definition of the P\'{e}clet number slightly differs in the two studies.  However, in both cases the P\'{e}clet number varies across the transition range by about two orders of magnitude (Fig.~\ref{F:f9}(b)).

We now discuss the transition states shown in Figs.~\ref{F:f3}(d,e) that consist of two clearly distinguished regions with different patterns. They occur at intermediate P\'{e}clet numbers or, more general, in the transition region between deposition of continuous states (Figs.~\ref{F:f3}(a,c)) and of non-continuous patterns (Figs.~\ref{F:f3}(f) to (i)). As the contact line moves away from the open end of the chamber we observe in both cases in Figs.~\ref{F:f3}(d,e) a rather sharp transition from the deposition of stripes to the deposition of a continuous film. This can be explained as caused by the continuous drift in the P\'{e}clet number. As indicated in Fig.~\ref{F:f9}(a), the P\'{e}clet number decreases with the increase of the distance between the open end of the chamber and the moving contact line, $x$, due to the decrease in the evaporation rate of the solvent. The latter results from the  continuously increasing path length that the toluene vapor has to travel from the meniscus to the ambient air outside the chamber. A careful choice of experimental parameters allows us to commence our measurement with a large P\'{e}clet number at small $x$ and end the experiment with a much smaller P\'{e}clet number at large $x$ values (Fig.~\ref{F:f9}). The difference can be up to approximately two orders of magnitudes. This implies that the pattern can ``drift'' in the course of a single experimental procedure, changing from an non-continuous pattern deposition, found for large P\'{e}clet numbers, to continuous deposition, found for low P\'{e}clet numbers; see Figs.~\ref{F:f3}(d) and (e).

Finally, we conclude this part of the discussion by noting that although, overall we find some good qualitative agreement between the modeling results of the study by~\citet{73Fra} and our experimental results, there are also some points that call for cautiousness and require future improvements in model development and analysis:

(i) Even the most refined models are still rather restricted in the physical effects they allow for. Features that should be incorporated into the hydrodynamic models are the dependence of wettability on solute concentration, a distinction of dissolved solute and deposited solute (similar to the work conducted by \citet{OkKD2009pre}) and the possibility of phase-transitions of the solute-solvent mixture (as in \cite{NaTh2010n}). All such additions need to be incorporated with care as potential cross-couplings can be easily missed. Recently, a general theoretical framework has been proposed for thin-films of complex fluids based on a gradient dynamics approach (see \cite{ThAP2016prf,Thie2018csa} and references therein) - the general approach has not yet been systematically applied to models for deposition from solution although several have been brought into the corresponding gradient form \cite{WTGK2015mmnp}.

(ii) The second issue concerns the analysis of the experimental patterns. Here we have determined pattern types and discussed their main features but have not given any further quantitative measures. In contrast, some theoretical works, based on time simulations, have analyzed the dependency of properties of line patterns on control parameters \cite{KGFT2012njp,DeDG2016epje}. In particular, the onset of pattern formation is of interest. Hypothesis regarding the underlying bifurcations are formulated elswhere \cite{73Fra} based on time simulations, while a different study \cite{KoTh2014n} was able to employ continuation techniques to determine the full bifurcation diagram for line deposition in Langmuir-Blodgett transfer. Here, we see much future scope for detailed experimental analyses. 

(iii) The final point concerns a major challenge faced by the theoretical approach. Nearly all mentioned theoretical studies of deposition patterns (and dip coating and Langmuir-Blodgett transfer patterns) are limited to a one-dimensional (1d) substrate implying that the models can only distinguish between regularly spaced and irregularly spaced deposits of straight lines, and do not provide detailed information on transitions between the different two-dimensional (2d) pattern variants described here. There are a few exceptions in studies of the evaporative dewetting of nanoparticle suspensions \citep{VTPS2008pre,RoAT2011jpcm}, Langmuir-Blodgett transfer of surfactant layers \citep{LKGF2012s,KGFT2012njp,WTGK2015mmnp}, and the dip-coating of a substrate in solutions \citep{WTGK2015mmnp,EGUW2018springer} that do, however, not provide an overall picture in terms of a bifurcation analysis.


\subsection{The oscillatory wetting-dewetting motion}
Most of the previous experimental and theoretical studies in the literature describe the motion of the contact line during the process of pattern deposition as a stick-slip dynamics or as consisting of a sequence of pinning and depinning processes, which are connected to capillary forces~\cite{thompson,hoffman}, the geometry of the deposits~\cite{Bodiguel:2010hh}, or viscous forces~\cite{73Fra}. Here, however, we observe a different type of contact line dynamics, namely, cycles of oscillatory wetting-dewetting motions of the contact line (Fig.~\ref{F:f4}). In each cycle the displacement during the dewetting phase is larger than the one in the wetting  phase such that, overall, the liquid front recedes in time. The oscillatory motion of the contact line is likely the result of an additional effect, which is related to variations in the surface tension of the solution near the contact line. The variations in the surface tension appear to result as a consequence of a concentration-dependence of the surface tension, i.e., a Marangoni effect. Such effects were discussed previously \cite{Doumenc:2013bh} but not in the context of a non-monotonic contact line motion, which is observed here.

In the case that the polymer possesses a lower surface tension than the solvent (in contact with air), the rapid increase in the concentration of the polymer near the contact line results in a decrease of the local surface tension of the solution close to the contact line. The resulting Marangoni forces are directed away from the contact line into the meniscus and cause a flow that pushes liquid from the vicinity of the contact line into the meniscus \cite{poulard2007,26AboJabal}. In the opposite case, if the polymer possesses a higher surface tension than the solvent, a fast increase in the polymer concentration near the contact line causes a sharp local increase in the surface tension of the solution. The resulting Marangoni flow then pushes liquid from the bulk of the meniscus into the vicinity of the contact line. If the resulting flux is fast, the direction of contact line motion may reverse, resulting in the wetting phase of the cycle \cite{poulard2007,29Fanton,pesach}. In our case, the surface tension of toluene at $20^\circ$C is approximately 28.4 mN/m, while the surface tension of PMMA at the same conditions is approximately 40 mN/m \cite{BKbrandrup1975polymer}. Hence, one may expect that the accumulation of PMMA near the contact line increases the local surface tension and may trigger the observed phase of dynamic wetting.

%

\begin{figure}[H]
    \centering
  \includegraphics[width=3.25 in]{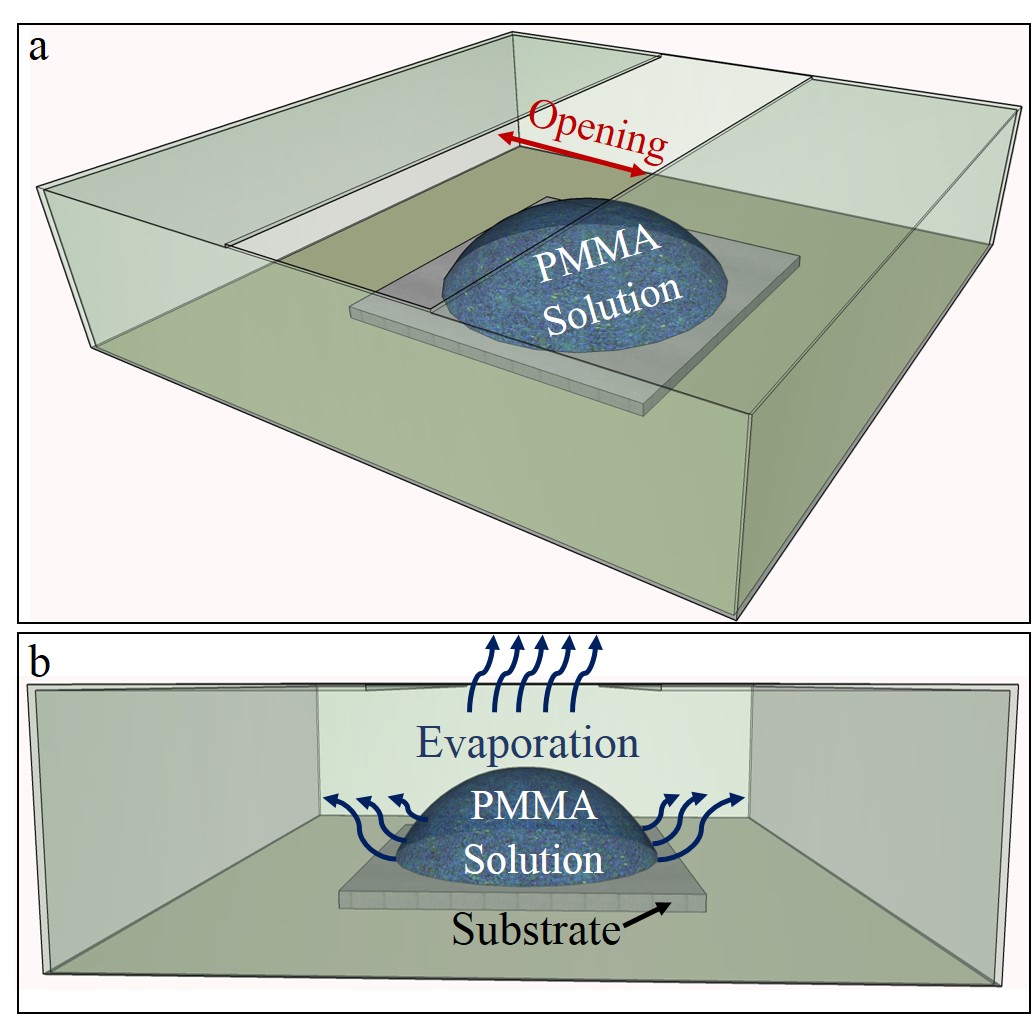}\\
  \caption{Sketches of the (a) top view and (b) side view of the semi-closed void system used to qualitatively control the rate of evaporation for a sessile drop of a volatile polymer solution.}
  \label{F:f42}
\end{figure}

\begin{figure}[H]
    \centering
  \includegraphics[width=6.5 in]{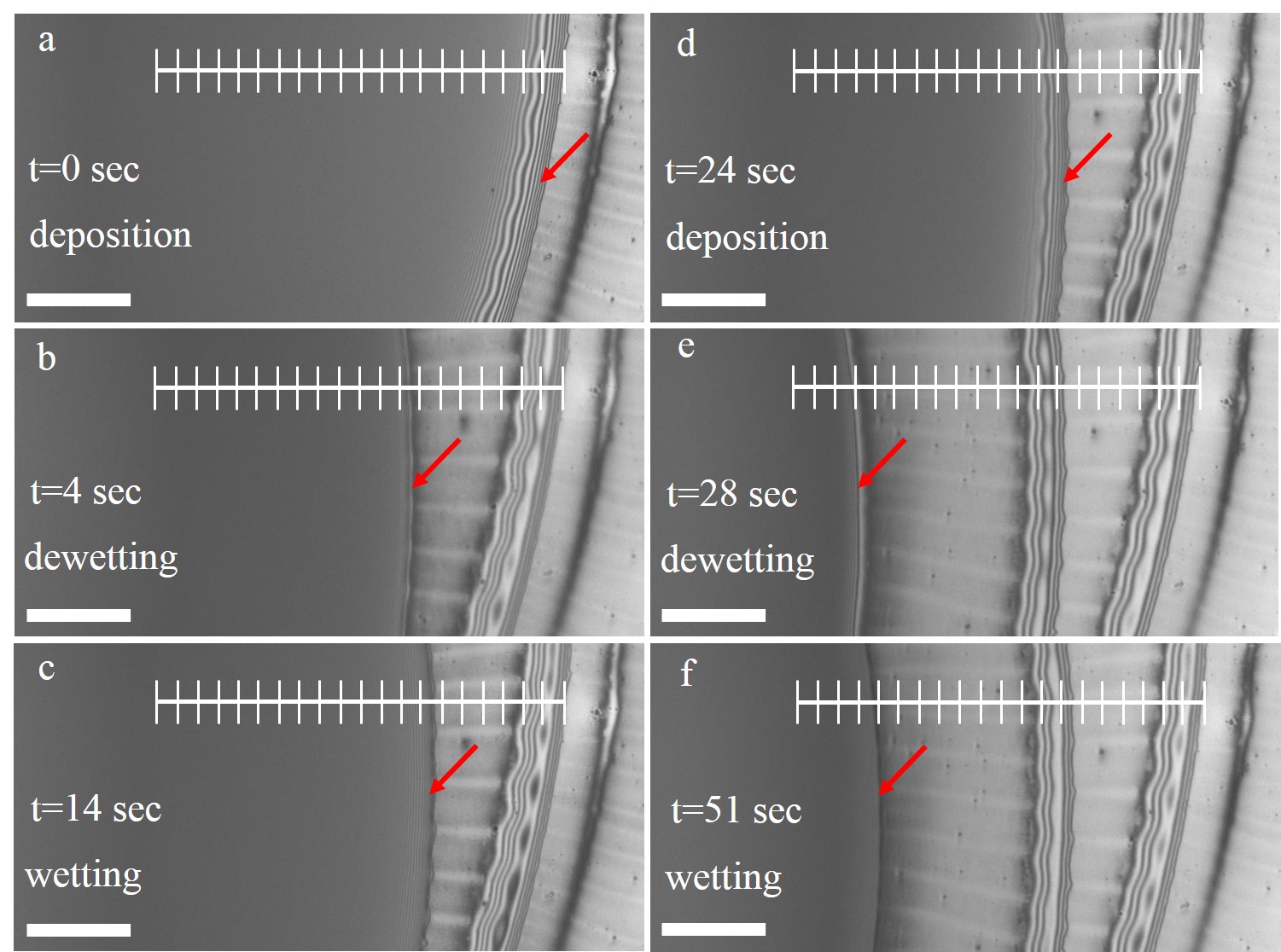}\\
  \caption{The oscillatory wetting-dewetting motion of the three phase contact line of a drop inside a cavity, which is partially open to the atmosphere, is depicted as a sequence of optical microscope frames. Frames (a-c) and (d-f) respectively represent two subsequent cycles. The contact line (marked by a red arrow) separates the solution on the left (appearing as a nearly uniform patch of gray) and the polymer deposit on the right. Both, the polymer deposit and the liquid in the vicinity of the contact line appear as dark and bright fringes due to the diffraction of light. The temperature is $T=25^{\mathrm{o}}{\mathrm{C}}$, the initial polymer (PMMA) concentration is $C=1$ mg/ml, and $M_w$=350,000 Da. The scale bars shows 100 $\mu m$ and the tick separation  is 20 $\mu$m.  } 
  \label{F:f41}
\end{figure}

To verify that the oscillatory motion of the contact line is not specific to the micro-chamber geometry, we also conduct a similar experiment employing a drop of a volatile polymer solution. The drop on a silicon oxide substrate, is placed at the bottom of a void composed of a glass box whose upper side has an adjustable opening as illustrated in Fig.~\ref{F:f42}. By adjusting the width of the opening, the contact area to the ambient air is controlled. This allows us to qualitatively alter the rate of evaporation of the drop. With this setup we also find the described oscillatory wetting-dewetting motion of the contact line, as illustrated by the snapshots in Fig.~\ref{F:f41} and by the movie SI3 in the Supporting Information. Decreasing the evaporation rate by decreasing the size of the opening, we observe a transition in the contact line motion from stick-slip motion to oscillatory wetting-dewetting motion. The sequence of frames in the figure shows two cycles of the latter with  the contact line marked by the red arrow. Following the deposition of the stripe (shown in (a) as a light diffraction pattern), which is placed parallel to and on the right side of the contact line, the contact line recedes and dewets the substrate until its motion is arrested in (b). Then, the contact line moves again but in the opposite direction wetting the substrate again, as shown in (c). Images (d) to (f) present the next dewetting-wetting cycle, where the deposition of a new stripe takes place in (d). The solution dewets the substrate until being arrested again in (e). Then, the motion of the contact line again changes direction and the solution wets the substrate. The next stripe deposition occurs at the position of the contact line in (f).

%

\par

\par

\section{Conclusions}\label{conclusion}
We  have studied the evaporative dewetting of a volatile polymer solution and the patterns of polymeric deposit left behind by the moving three-phase contact line. Two geometries, namely, a meniscus in a micro-chamber and a sessile drop in a semi-closed void have consistently shown the occurrence of different modi of contact line motion and deposition depending on initial polymer concentration, molecular mass of the polymer, and ambient temperature.

We have shown that the motion of the meniscus is governed by a P\'{e}clet number, characterizing the ratio between the convection of the solution and the diffusion of the solvent. It is also proportional to the rate of liquid evaporation. A small P\'{e}clet number results in diffusion-dominated transport of polymer chains, which causes a nearly homogeneous solute concentration in the solution and a monotonic contact line motion. This in turn gives either a continuous deposition of a nearly uniform polymer film or a continuous deposition that does not dry as a uniform solid polymer film but undergoes a further continuous dewetting step, resulting in a dry polymer film which contains a isotropic arrangement of many small holes or a random continuous polygonal network. 

In contrast, at large P\'{e}clet numbers, polymer diffusion is very slow as compared to the evaporation-induced convective flux. Ultimately, this convection-dominated transport results in the deposition of various types of non-continuous patterns, namely, regular sequences of parallel lines, line patterns with side protrusions caused by transversal contact line instabilities, and 
``punch-hole-like'' patterns caused by strong transversal contact line instabilities. 
Using morphological phase diagrams we have summarized how the type of deposit depends on temperature and initial concentration. Notably, in this case the contact line may show either the usual stick-slip motion where in the slip phase the contact line recedes, i.e.,  dewetting of the solid, or an oscillatory wetting-dewetting motion where the contact line ratchets back and forth, on average resulting in a receding motion. Both motion types support deposits in the form of stripes, and further may show fingering instabilities during the dewetting phase when increasing the temperature and/or concentration in the cases of higher molecular weight.   

We have explained the mechanisms responsible for the oscillatory wetting-dewetting motion as follows: The evaporation of the solvent supports the accumulation of the polymeric solute close to the contact line. This increase in solute concentration increases the local surface tension and the resulting Marangoni flow is directed towards the contact line and supports the dynamic wetting phase. Eventually, the continuously ongoing evaporation of the solvent solidifies (diverging viscosity at gelling/jamming transition) the solution near the contact line to a level where the motion stops. Then the solution away from the contact line breaks from the solidified part and undergoes a fast convective dewetting before solvent evaporation dominates again. When the concentration of the solute near the contact line is again sufficiently increased to alter the direction of motion of the contact line, again the dynamic wetting phase occurs and the cycle starts from the beginning.

 A qualitative control over the rate of solvent evaporation from a drop of the same polymer solution in a void has further shown that the wetting-dewetting motion cycles are not limited to the micro-chamber geometry but occur in different confined geometries. In both considered geometries one may shift from a stick-slip motion of the contact line to an oscillatory wetting-dewetting motion by altering the evaporation rate.


We have compared our experimental results to theoretical results, which are available in the literature. The theory predicts the deposition of periodic lines in several similar settings. We find qualitative agreement in several major points. Most important of which is the connection between the morphology of the deposit and the ratio between the rate of convection and diffusion of the polymer chains in the volatile solvent. However, we have also pointed out in which way the existing models are limited and explained which additional effects should be included in future theoretical models.

\section*{Acknowledgements}
We acknowledge support of this research by the German Israel Foundation for Scientific Research and Development (GIF) under grant number I-1361-401.10/2016.

\bibliography{reference}
\bibliographystyle{unsrt}
\end{document}